\documentclass[lettersize,journal]{IEEEtran}
\usepackage{amsmath,amsfonts}
\usepackage{algorithmic}
\usepackage{array}
\usepackage[caption=false,font=normalsize,labelfont=sf,textfont=sf]{subfig}
\usepackage{textcomp}
\usepackage{stfloats}
\usepackage{url}
\usepackage{verbatim}
\usepackage{graphicx}
\usepackage{bm}
\usepackage{cite}
\usepackage{booktabs}
\usepackage{tcolorbox} 

\def\BibTeX{{\rm B\kern-.05em{\sc i\kern-.025em b}\kern-.08em
    T\kern-.1667em\lower.7ex\hbox{E}\kern-.125emX}}
\usepackage{balance}
\begin{document}
\bstctlcite{BSTcontrol}
\title{Edge-Native Embodied Intelligence for Action-Aware Wireless Edge Networks}

\author{Yiru Wang, Chuanao Jiang, Jiahui Cui, Zide Fan,~\IEEEmembership{Member,~IEEE, }Lei Wang,~\IEEEmembership{Member,~IEEE, }

Zehui Xiong,~\IEEEmembership{Senior~Member,~IEEE, } Dong In Kim,~\IEEEmembership{Life Fellow,~IEEE }

\thanks{Yiru Wang, Chuanao Jiang, Zide Fan and Lei Wang are with the Aerospace Information Research Institute, Chinese Academy of Sciences, Beijing 100190, China, and also with the Key Laboratory of Target Cognition and Application Technology, Aerospace Information Research Institute, Chinese Academy of Sciences, Beijing, China (wangyiru@aircas.ac.cn; jiangcy@aircas.ac.cn; fanzd@aircas.ac.cn; wanglei002931@aircas.ac.cn);
Jiahui Cui is with the Aerospace Information Research Institute, Chinese Academy of Sciences, the Key Laboratory of Target Cognition and Application Technology, Aerospace Information Research Institute, Chinese Academy of Sciences, University of Chinese Academy of Sciences, and also with School of Electronic, Electrical and Communication Engineering, University of Chinese Academy of Sciences (cuijiahui24@mails.ucas.ac.cn);
Zehui Xiong is with the School of Electronics, Electrical Engineering and Computer Science (EEECS), Queen's University Belfast, Belfast, BT7 1NN, U.K. (z.xiong@qub.ac.uk); Dong In Kim is with the Department of Electrical and Computer Engineering, Sungkyunkwan University, Suwon 16419, South Korea (dongin@skku.edu).
}}

\markboth{Journal of \LaTeX\ Class Files,~Vol.~18, No.~9, September~2020}%
{How to Use the IEEEtran \LaTeX \ Templates}

\maketitle

\begin{abstract}

Embodied intelligence is shifting artificial intelligence from passive digital perception toward active physical interaction. However, foundation-model-enabled embodied agents face a fundamental tension between open-world cognition and
resource-constrained deployment. On-device models are limited by computation, memory, and energy budgets, whereas cloud-centric solutions introduce latency and reliability risks over dynamic wireless links. Edge general intelligence provides
a promising cognitive backbone, but existing frameworks still lack physical grounding, action awareness, and mechanisms for actively acquiring useful physical experience. To address these limitations, this article introduces edge-native embodied intelligence (ENEI), an action-aware wireless edge framework that integrates embodied agents, the 6G communication and networking fabric, and edge cognitive services into a 6G-mediated bidirectional edge-embodiment loop. Along the edge-to-embodiment axis, confidence-aware assistance and edge-driven generative adaptation enhance local autonomy under out-of-distribution (OOD) conditions. Along the embodiment-to-edge axis, value-of-experience guided active embodied federated learning enables physical actions to generate informative experience for continuous edge model evolution. The 6G fabric supports both directions through goal-oriented transmission and programmable radio-resource allocation. Two case studies on OOD drone navigation and mobility-driven federated learning illustrate the feasibility and communication efficiency of the proposed mechanisms. ENEI provides a unified perspective in which edge cognition strengthens embodied action, while embodied agency actively enriches edge cognition, laying the foundation for scalable, adaptive, and self-evolving embodied wireless systems.

\end{abstract}


\section{Introduction}
\IEEEPARstart{E}{mbodied} intelligence is becoming a key frontier of artificial intelligence (AI), shifting intelligent systems from passive digital perception toward active physical interaction. Rooted in cybernetics and developmental psychology, embodied intelligence assumes that an intelligent agent is situated in an environment and learns through a continuous perception-action-interaction loop~\cite{liu2025aligning}. This embodied paradigm has recently been further advanced by foundation models, including large language models (LLMs)~\cite{black2025pi05} and vision-language-action (VLA) models~\cite{zitkovich2023rt}, which provide embodied agents with stronger general-purpose cognitive capabilities. By leveraging broad pre-training and multimodal reasoning, these models can interpret high-level goals, reason over semantic context, and generate executable action plans. However, their large parameter scales and intensive computation make direct deployment on battery-powered embodied devices difficult, especially under strict latency, energy, and size constraints.

Existing deployment paradigms only partially address this tension. Cloud-based offloading can provide abundant computation, but long-haul transmission and unstable wireless links introduce latency, jitter, and reliability risks that are unacceptable for safety-critical physical actions~\cite{Chen2024Edge}. On-device compression techniques, such as knowledge distillation and model pruning, reduce inference latency~\cite{zhu2025minivln}, but they also constrain model capacity and weaken open-world adaptability. Edge general intelligence (EGI) offers a more promising middle ground by moving general-purpose inference, memory, and model services closer to end devices~\cite{zhang2026toward}. With proximity to users and stronger computation than local devices, EGI can serve as a cognitive backbone for embodied systems.

Nevertheless, EGI alone does not complete the embodied intelligence loop. Current edge intelligence frameworks are still largely organized around digital data processing and service provisioning. They lack physical embodiment, direct environmental feedback, and action-level awareness. As a result, the edge may allocate scarce communication and computation resources according to data volume, semantic fidelity, or channel quality, while overlooking whether the transmitted information can actually change an agent's physical behavior or reduce operational risk. More importantly, without embodied agents actively exploring the world, edge models remain vulnerable to cognitive blind spots caused by passive, biased, and incomplete data collection. This limitation motivates a new architecture in which edge intelligence and embodied physical agency are not treated as separate entities, but as mutually reinforcing components of a wireless edge ecosystem.

Recent studies have approached this convergence from different perspectives. Existing edge intelligence studies mainly enhance wireless edge systems in the digital domain through semantic communication, distributed inference, and agentic autonomy~\cite{yang2022semantic,zhang2026toward}. In parallel, 6G-enabled embodied intelligence studies investigate the integration of wireless networks and embodied agents, emphasizing perception-communication-action coupling, wireless-embodied synergy, and communication-control co-design~\cite{embodied_6g,wireless_embodied,b2x}. Another emerging direction rethinks wireless systems or agentic communication as adaptive, environment-aware, or reasoning-aligned entities~\cite{eiw,reasoning_native_6g}. Despite these advances, most existing paradigms still emphasize one dominant axis, such as intelligent networking, wireless support for embodied agents, or embodied wireless systems. They do not fully organize edge cognition, 6G interaction, and embodied physical agency into a unified and co-evolving architecture.

\begin{figure*}
	\centering
	\includegraphics[width=18cm]{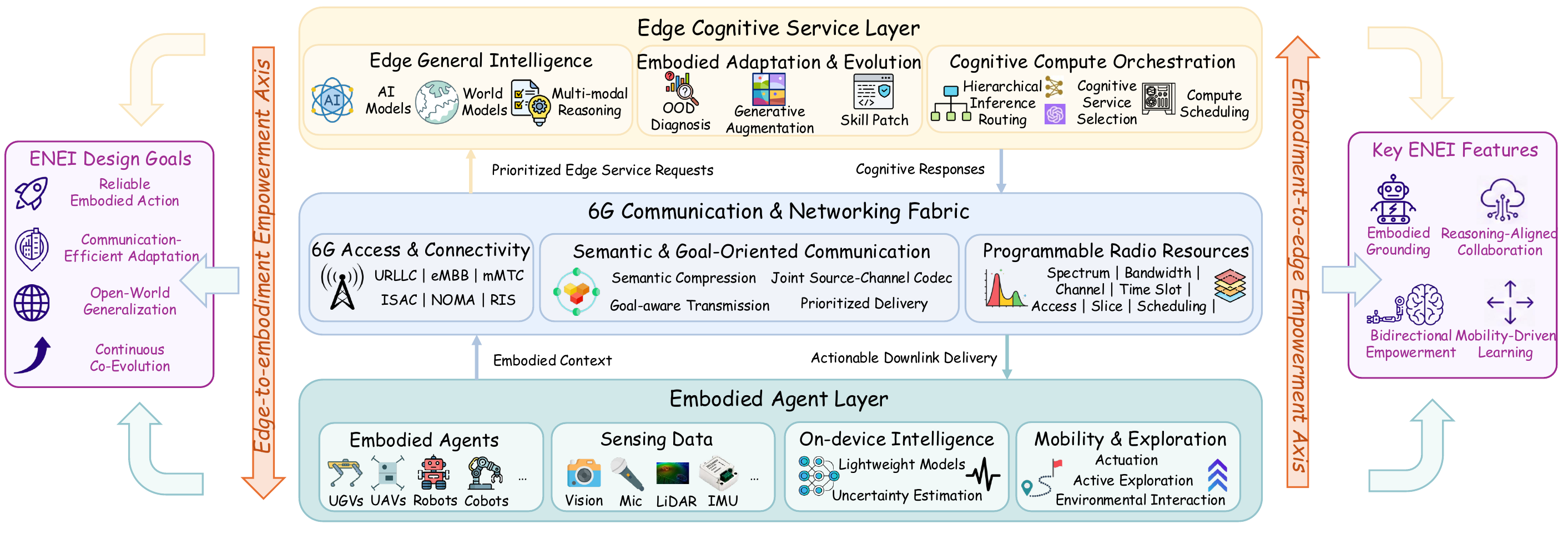}
	\caption{The ENEI architecture for action-aware wireless edge networks. Embodied agents perceive the physical environment, execute task-driven actions, and generate action-grounded context and experience. The 6G communication and networking fabric performs action-aware context engineering and programmable wireless delivery, while edge cognitive services provide collaborative reasoning, embodied adaptation, and cognitive compute orchestration. Together, the three layers form a bidirectional edge-6G-embodiment loop, where edge cognition supports coordinated physical actions and embodied interaction continuously drives edge intelligence evolution.}
	\label{fig:architecture}
\end{figure*}


To fill this gap, this article proposes Edge-Native Embodied Intelligence (ENEI), an edge-6G-embodiment architecture for action-aware wireless edge networks, as illustrated in Fig.~\ref{fig:architecture}. Different from existing related frameworks, ENEI places physical action at the center of the edge-embodiment interaction. From edge to embodiment, edge cognition is provided as action-aware cognitive services that support safety-relevant and task-critical physical decisions. From embodiment to edge, physical actions and environmental interactions generate action-grounded feedback and experience that continuously refine edge intelligence. The 6G communication and networking fabric serves as the programmable interaction medium between the two sides, engineering action-relevant context into compact and prioritized cognitive requests, orchestrating wireless resources according to task and inference requirements, and delivering edge cognition back as timely and actionable support. Together, these interactions form a 6G-mediated bidirectional edge-embodiment loop that supports continuous ENEI evolution. Table~\ref{tab:comparison} further compares ENEI with representative related frameworks.

The main contributions of this article are summarized as follows:
\begin{itemize}
    \item We propose ENEI, an action-aware edge-6G-embodiment architecture that integrates embodied agents, the 6G communication fabric, and edge cognitive services into a unified bidirectional framework.

    \item We develop the edge-to-embodiment axis through confidence-aware offloading and edge-driven generative adaptation, validated by a drone-navigation case study under environmental degradation.

    \item We develop the embodiment-to-edge axis through value of experience (VoE)-guided active embodied federated learning (AEFL), where embodied mobility actively acquires valuable experience for edge model evolution, validated by a moving-agent federated learning case study.

\end{itemize}

\section{ENEI Architecture for Action-Aware 6G Wireless Edge Networks}
\label{sec:architecture}

This section presents the three-layer ENEI architecture and its four information flows.

\begin{table*}[t]
    \centering
    \caption{Comparison between ENEI and existing related frameworks.}
    \label{tab:comparison}
    \includegraphics[width=18cm]{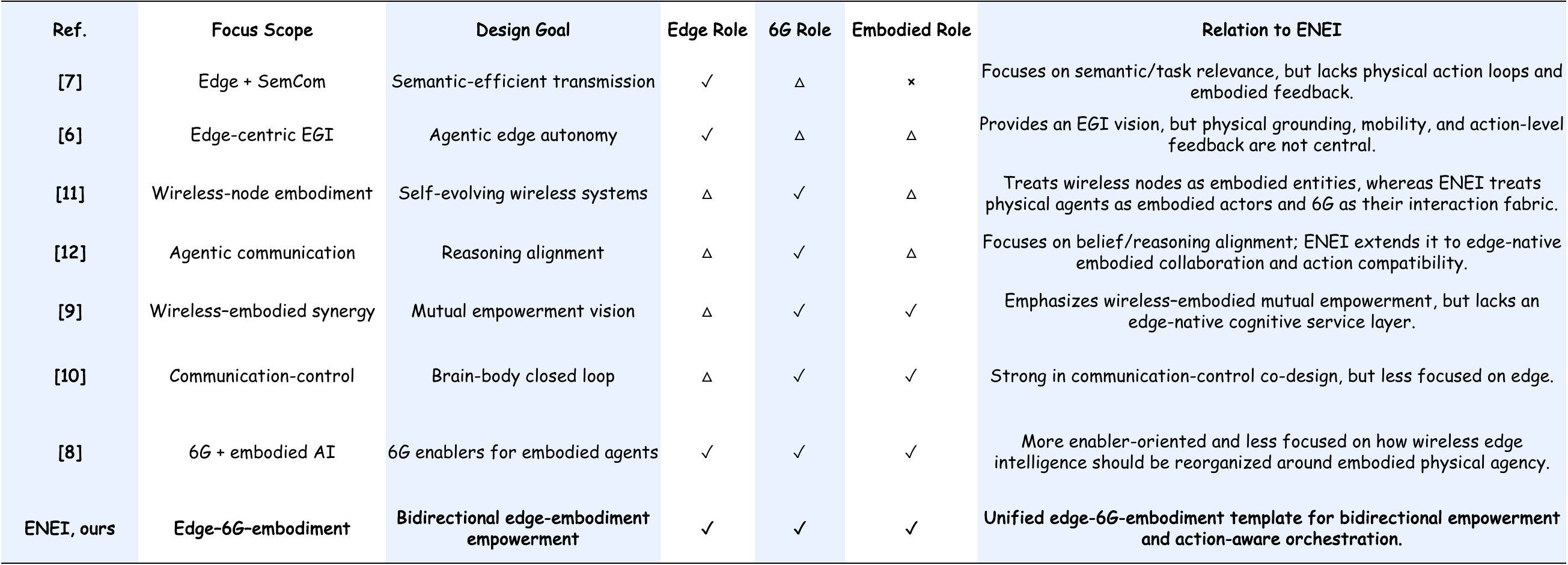}
\end{table*}

\subsection{Core Architectural Components}

\subsubsection{Embodied Agent Layer}

The embodied agent layer serves as the physical entry point of ENEI. 
It consists of heterogeneous agents, such as unmanned ground vehicles, unmanned aerial vehicles, and robots, equipped with multimodal sensors to perceive and interact with dynamic environments. These agents collect sensory observations, transform them into task-related local states, and run lightweight on-device intelligence for perception, uncertainty estimation, and fast decision support.

Different from conventional mobile devices, embodied agents possess closed-loop physical agency. Their sensing behavior is shaped by motion and exploration, while their future actions are influenced by current perception, local uncertainty, and task goals. 
Therefore, the information generated by this layer is naturally action-conditioned, motivating the transmission of embodied context rather than raw sensory data alone.

\subsubsection{6G Communication and Networking Fabric}

The 6G communication and networking fabric is the action-aware networking interface between embodied agents and edge cognitive services. 
It provides heterogeneous wireless access to support large-scale embodied agents in dynamic physical environments. 

Beyond connectivity, the 6G fabric performs semantic and goal-oriented transmission and supports inference-budget-aware context engineering. Wireless and embodied context is selectively structured and compressed according to edge inference needs, jointly considering expected action impact, communication cost, and inference overhead. This enables limited edge inference capacity to focus on decision-critical information while meeting stringent latency requirements.

The fabric further allocates programmable radio resources, including spectrum, bandwidth, channels, time slots, access modes, and network slices, according to payload, latency, reliability, and service-priority requirements. Its main role is therefore to transform heterogeneous agent-side context into compact and prioritized edge service requests and deliver edge-generated cognitive outputs as actionable downlink support.



\subsubsection{Edge Cognitive Service Layer}

Beyond goal-oriented semantic communication, semantic correctness or task relevance alone does not necessarily guarantee coordinated physical execution. The same semantically correct message may induce different, and potentially conflicting, actions when embodied agents differ in their physical states, beliefs, roles, or local decision policies. Edge general intelligence addresses this gap through collaborative reasoning over heterogeneous agent states. It integrates AI models, world models, and multi-modal reasoning modules to build a shared task-level understanding, reason about how different agents may respond to common information, and generate agent-specific cognitive outputs, such as sub-goals, coordination constraints, or action priors. Rather than enforcing identical actions, these outputs steer heterogeneous local policies toward mutually compatible behaviors under shared task and safety requirements.

Beyond collaborative reasoning during task execution, embodied agents may experience persistent capability degradation as environments and tasks evolve. The embodied adaptation service addresses this longer-term challenge by diagnosing recurring local failures, generating task-relevant adaptation data, and distilling the acquired knowledge into compact skill patches. These updates progressively enhance local models and reduce repeated dependence on edge assistance under changing operating conditions.

Cognitive compute orchestration manages heterogeneous cognitive requests under constrained device and edge inference budgets. Embodied agents may encounter tasks with different action urgency and reasoning complexity, making uniform inference provisioning inefficient. Inspired by hierarchical collaboration~\cite{dong2026aerial}, ENEI coordinates inference across embodied devices and the edge: lightweight and time-sensitive reasoning can remain on-device, while more complex requests are escalated to edge cognitive services. At the edge, requests can be further assigned to specialized reasoning modules or more capable foundation models according to service requirements and available computing resources. The orchestrator jointly coordinates inference placement, service selection, scheduling, and compute allocation across concurrent requests, enabling heterogeneous agents to receive appropriate cognitive support without indiscriminately invoking resource-intensive models.

\subsection{Information Flow in ENEI}

The proposed ENEI architecture is driven by four major information flows.

\subsubsection{Embodied Context Upload}

The first flow is from the embodied agent layer to the 6G fabric. Embodied agents generate embodied context, which may include sensory observations, compressed features, agent states, and environmental interaction feedback. Compared with raw data packets, embodied context is action-conditioned. It describes not only what the agent observes, but also what the agent may do next and what physical risk may arise.

\subsubsection{Prioritized Edge Service Requests}


The second flow is from the 6G fabric to the edge cognitive service layer. The 6G fabric selectively structures and compresses wireless and embodied context according to edge inference needs, expected action impact, and communication and inference costs.
The resulting service requests contain compact task-relevant context, priority information, and requested cognitive services, such as reasoning, adaptation, or retrieval.

\subsubsection{Cognitive Responses}

The third flow is from the edge cognitive service layer to the 6G fabric. Based on the received service requests, the edge
generates agent-specific cognitive outputs such as high-level guidance, sub-goals, or skill patches. It also attaches delivery requirements, including payload size, latency deadline, reliability target, and learning-value indicators.

\subsubsection{Actionable Downlink Delivery}

The fourth flow is from the 6G fabric to the embodied agent layer. The network delivers edge-generated, agent-specific outputs as actionable downlink information under latency, reliability, bandwidth, and task-urgency constraints. Although different agents may receive different guidance and consequently take different physical actions, these actions are expected to remain compatible with shared task objectives and coordination constraints. The 6G fabric therefore does not merely forward edge outputs, but supports their timely and reliable delivery for coordinated physical execution.


\begin{tcolorbox}[title=Wireless Design Lessons from ENEI]
\textbf{1) From data packets to embodied context:} Uplink transmission should carry task goals, uncertainty, and action risk, not only sensory data. 

\textbf{2) From semantic relevance to action impact:} Scheduling should prioritize information that can change physical decisions before the action deadline.


\textbf{3) From static quality-of-service (QoS) to action-aware QoS:} Latency, reliability, and bandwidth should be configured according to physical risk, task urgency, and model-update value.

\textbf{4) From context abundance to inference-efficient context:}
Under limited edge inference budgets, the 6G fabric should retain context with high expected action impact, reducing inference latency while preserving decision-critical information.
\end{tcolorbox}

\section{Edge-to-Embodiment Cognitive Empowerment}

\begin{figure*}
	\centering
	\includegraphics[width=18cm]{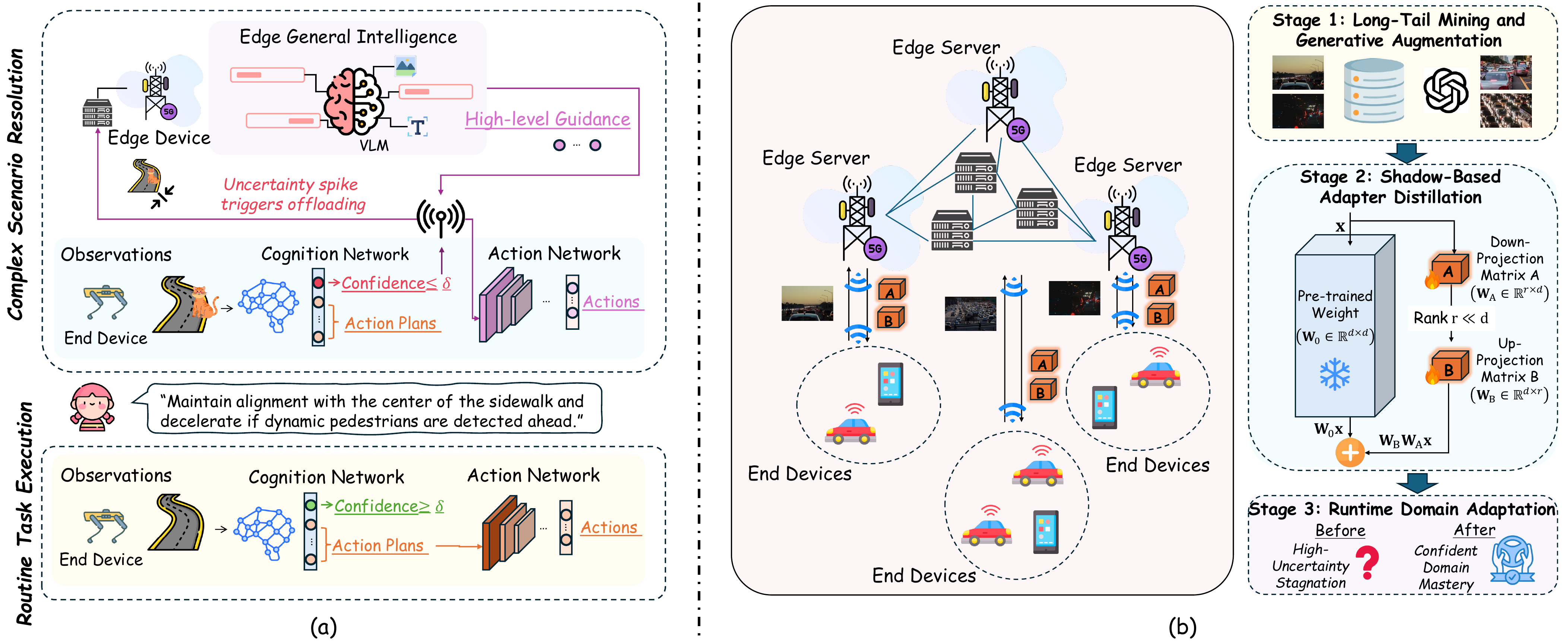}
	\caption{The edge-to-embodiment cognitive empowerment axis. (a) Confidence-aware adaptive offloading: embodied agents handle routine states locally and invoke edge reasoning only when action-relevant confidence degradation may affect safety-critical or task-critical decisions. (b) Edge-driven generative adaptation: the edge uses a shadow model to diagnose OOD failures, synthesize task-relevant adaptation data, and distill compact skill patches for wireless delivery.  Together, these mechanisms enable edge cognitive services to assist resource-constrained embodied agents without continuously offloading every perception--action loop.}
	\label{fig:edge for ei}
\end{figure*}

The edge-to-embodiment axis aims to reconcile real-time local control with open-world generalization. This tension arises because the two capabilities are typically provided by different levels of intelligence: foundation models offer strong reasoning and generalization but incur substantial inference and communication overhead, whereas lightweight onboard models support fast local execution but remain vulnerable to unfamiliar or long-tail conditions.

ENEI addresses this tension by combining confidence-aware adaptive offloading with edge-driven generative adaptation. The former invokes edge reasoning only when uncertainty may affect an imminent action; the latter delivers compact skill patches that restore the local perception-action loop under out-of-distribution (OOD) conditions. Through the 6G fabric, prioritized embodied context is uploaded and cognitive guidance or skill updates are returned as actionable downlink support.

\subsection{Confidence-Aware Adaptive Offloading}

Confidence-aware adaptive offloading addresses the cognition--latency dilemma by determining when an embodied agent should rely on local inference and when it should request edge assistance. Unlike static offloading policies that are mainly triggered by channel quality or computing load, this mechanism uses action-conditioned predictive uncertainty as a decision signal. The key intuition is that familiar and low-risk states should be handled locally, while confidence degradation should trigger edge assistance only when it may materially affect imminent physical actions, safety constraints, or task progress.

As illustrated in Fig.~\ref{fig:edge for ei} (a), the embodied agent first processes sensory observations with its onboard model and estimates predictive uncertainty through indicators such as entropy or confidence scores. When the uncertainty remains below a predefined threshold, the agent directly executes the local policy. 
When an uncertainty spike occurs, the agent packages the relevant embodied context, such as semantic features, task descriptors, agent states, or failure snapshots, and sends a prioritized edge service request through the 6G fabric. Meanwhile, the agent can maintain a safe fallback behavior, such as slowing down, hovering, or switching to a conservative policy.

After receiving the request, the edge cognitive service layer performs deeper reasoning over the embodied context and returns actionable cognitive support. 
The response may take the form of high-level guidance, symbolic sub-goals, waypoint suggestions, latent policy hints, or lightweight model corrections, depending on the task requirement and downlink constraints. By combining local reflexive control with event-triggered edge reasoning, confidence-aware adaptive offloading avoids unnecessary wireless round trips for routine states, while preserving access to stronger edge cognition when local uncertainty may affect safety-relevant or task-critical physical actions.

\subsection{Edge-Driven Generative Adaptation}

Confidence-aware offloading provides on-demand access to edge reasoning, but it does not fully solve the long-term generalization problem of lightweight onboard models. 
When embodied agents repeatedly encounter OOD conditions, frequent edge intervention may increase communication load and weaken local autonomy. Therefore, the edge should not only serve as a temporary inference assistant, but also act as an adaptive knowledge provider that improves the agent's local capability.

To this end, we propose edge-driven generative adaptation. The key idea is to use the edge cognitive service layer to diagnose failures, synthesize task-relevant adaptation data, and distill the acquired knowledge into compact skill patches. Thus, it can restore the reliability of the local perception-planning-action loop so that the agent can make safe and timely navigation decisions without repeatedly invoking the edge.
Instead of transmitting full model checkpoints or continuously offloading every uncertain state, ENEI delivers lightweight skill updates through the 6G communication fabric, improving local robustness while keeping wireless payload and onboard computation affordable.

As demonstrated in Fig.~\ref{fig:edge for ei} (b), the mechanism can follow a shadow-based training paradigm, where the edge maintains a structural replica of the embodied device's model to train compatible adapters without interrupting local execution. When an embodied agent detects a long-tail scenario or repeated action uncertainty spike, it uploads a small number of failure cases and contextual observations. The edge analyzes these sparse physical samples to infer the underlying domain shift, and then uses foundation or generative models to construct a task-relevant adaptation set that preserves task semantics while covering the observed OOD factors.

Based on this adaptation set, the edge trains a lightweight adapter on the shadow model, such as a low-rank adaptation (LoRA) module~\cite{hu2022lora}. 
The resulting adapter captures scenario-specific knowledge as a compact parameter patch, which can be transmitted to the target agent or selectively shared with agents operating under similar conditions. 
After receiving the patch, the embodied device integrates it into the local cognition network, enabling the onboard model to better handle the encountered domain shift without replacing the full model.


\subsection{Case Study: Confidence-Aware Offloading and Edge Adaptation for OOD Embodied Drone Navigation}\label{Case Study}

\begin{figure*}
	\centering
	\includegraphics[width=18cm]{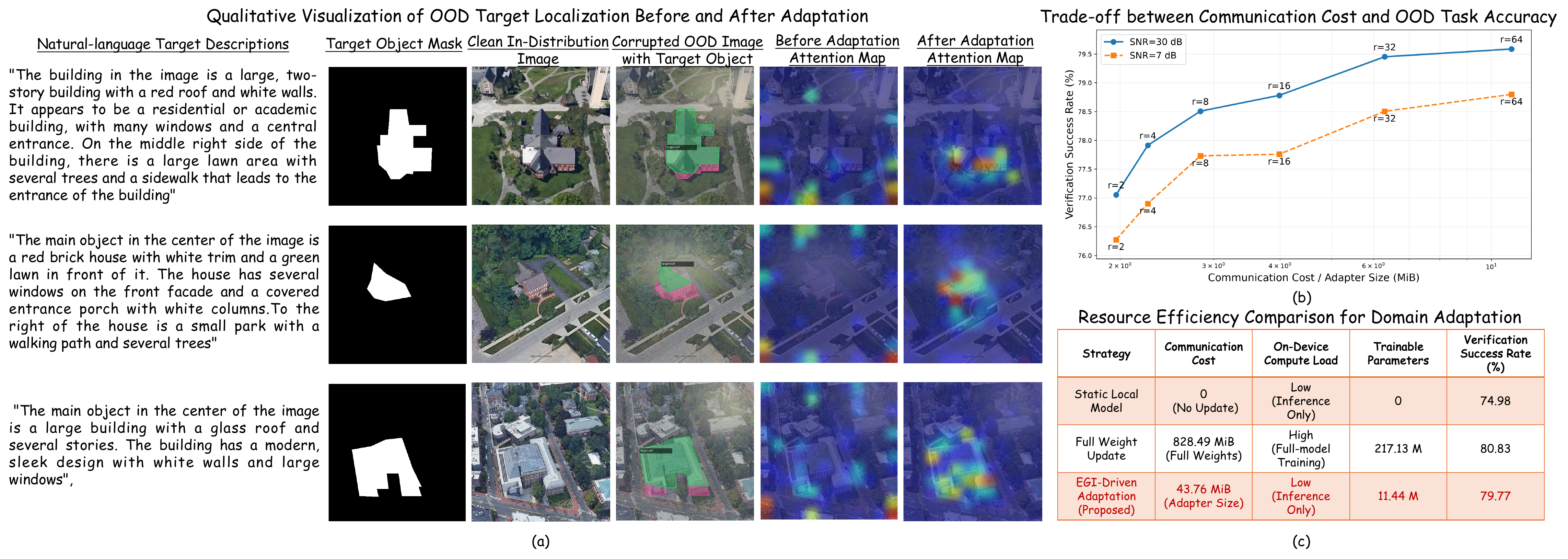}
	\caption{Performance evaluation of the proposed confidence-triggered edge adaptation under environmental shifts. (a) Action-relevant spatial grounding: target localization before and after adaptation, showing that the edge-generated skill patch restores attention to the visual regions required for navigation decisions. (b) Communication-action reliability trade-off: OOD verification performance versus adapter payload under different downlink SNR conditions, reflecting the wireless cost of recovering reliable local decision support. (c) Resource-efficient local autonomy: comparison of communication cost, onboard computation, trainable parameters, and verification performance, showing that compact edge adaptation approaches full-model updating while preserving lightweight local execution.}
	\label{fig:results}
\end{figure*}

To illustrate how edge intelligence can support reliable embodied action under environmental shifts, we consider natural-language-guided drone navigation using the real-world GeoText spatial-semantic dataset~\cite{chu2024towards}. In this task, the drone associates its current observation with a language-specified landmark before continuing along the planned route. The onboard verifier therefore provides the cognitive evidence required for the next navigation action. Pristine landmark images represent the nominal environment, while OOD observations are simulated by applying three representative visual degradations: over-exposure glare, dynamic shadow masking, and Gaussian sensor noise. These degradations emulate common perception failures in aerial navigation, where illumination changes, occlusions, and sensor noise can reduce the reliability of navigation decision.

We use the normalized entropy of the image-text matching output as a lightweight action-confidence indicator. Low entropy supports continued local execution, whereas an entropy spike indicates that the current perception may lead to an unreliable route decision. Under shifted operating conditions, the verification success rate persistently remains below 75\%. Thus, edge assistance is triggered by the potential effect of perception degradation on an imminent physical action.

Upon receiving the request, the edge employs Gemini-2.5-Pro\footnote{Google, ``Gemini 2.5 Pro,'' Google AI for Developers. Available: \url{https://ai.google.dev/gemini-api/docs/models/gemini-2.5-pro}, accessed Aug. 2026.} to characterize the observed degradation patterns and semantic attributes of the corrupted context, and then synthesizes a task-relevant adaptation set that mimics the OOD conditions while preserving the original landmark semantics. The set contains 3,366 unique images, including 2,866 edge-generated degraded samples and 500 clean counterparts. A lightweight LoRA adapter~\cite{hu2022lora} is then trained on an edge-side shadow replica of the drone verifier using a contrastive and region-aware objective that preserves image-text matching, strengthens target-region attention, and maintains consistency between clean and degraded observations. To evaluate the trade-off between wireless payload and adaptation performance, we vary the LoRA rank and compare the proposed edge-generated LoRA adaptation method against a static local model without update and a full-weight update baseline.

The resulting skill patch is intended not only to improve visual matching, but also to restore the target grounding and verification confidence required by the downstream navigation module. After deployment, similar shifted operating conditions can be handled more reliably by the onboard model, converting temporary edge assistance into persistent local capability.

To evaluate the wireless cost of this recovery, the adapter is quantized and transmitted as a compact differential update. We vary the LoRA rank and emulate noisy downlink transmission using a BPSK transmission over the additive white Gaussian noise channel under representative high- and low-SNR regimes, revealing the trade-off between adaptation payload and the reliability of the cognitive evidence available for subsequent actions.

As shown in Fig.~\ref{fig:results}(a), environmental shifts cause the onboard model to attend to dispersed background regions, whereas the adapted model re-concentrates its response on the language-specified target. On 32 building-stratified degraded images, the mean target-mask attention increases from 0.0751 to 0.3057, corresponding to a 4.07$\times$ improvement in localized semantic alignment.


Fig.~\ref{fig:results}(b) evaluates the wireless delivery of INT8-quantized adapters under different communication costs and channel conditions. The rank-64 INT8 adapter achieves 79.59\% verification success at 30~dB and remains effective at 78.16\% under 7~dB, indicating that most of the adaptation benefit is retained under communication-constrained delivery.

The resource comparison in Fig.~\ref{fig:results}(c) further confirms the efficiency of the proposed approach. A full-weight update improves OOD verification to 80.83\%, but requires an 828.49~MiB payload and updates 217.13 million parameters. The edge-generated adapter achieves 79.77\% with a 43.76~MiB payload and 11.44 million trainable parameters. This reduces both the trainable parameters and the model-update payload by 94.7\%, while approaching full-model performance with substantially lower communication and adaptation overhead.

Overall, this case study validates the edge-to-embodiment axis as an action-confidence restoration mechanism. When changing environments undermine the perceptual evidence required for a navigation action, the edge diagnoses the domain shift and delivers a compact skill patch that restores local target grounding and verification reliability. Hence, ENEI uses edge intelligence not merely to improve perception accuracy, but to recover the cognitive capability required for high-confidence local action. Closed-loop flight and trajectory-level evaluation remain important directions for future work.

\section{Embodiment-to-Edge Operational Orchestration}

The previous section considered how edge cognitive services improve embodied agents. The reciprocal direction is equally
important as embodied agents can improve edge intelligence through their sensing, mobility, and physical interaction. Conventional edge learning relies largely on passively available client data. Such data are often biased toward frequently visited environments and may omit rare but learnable physical situations. Uncertainty-based acquisition only partially addresses
this problem because high uncertainty does not necessarily imply high learning value. Sensor noise, glare, or intrinsically
ambiguous observations may remain uncertain even after repeated model updates.

In this regard, embodied agents provide a new degree of freedom. By changing their trajectories, sensing viewpoints, and interaction behaviors, they can reshape the future experience distribution rather than
merely selecting samples from a fixed local dataset. Based on this observation, ENEI introduces VoE-guided AEFL, in which
edge-side knowledge feedback guides physical experience acquisition, while the resulting federated updates further evolve
the edge model.

\begin{figure}
	\centering
	\includegraphics[width=9cm]{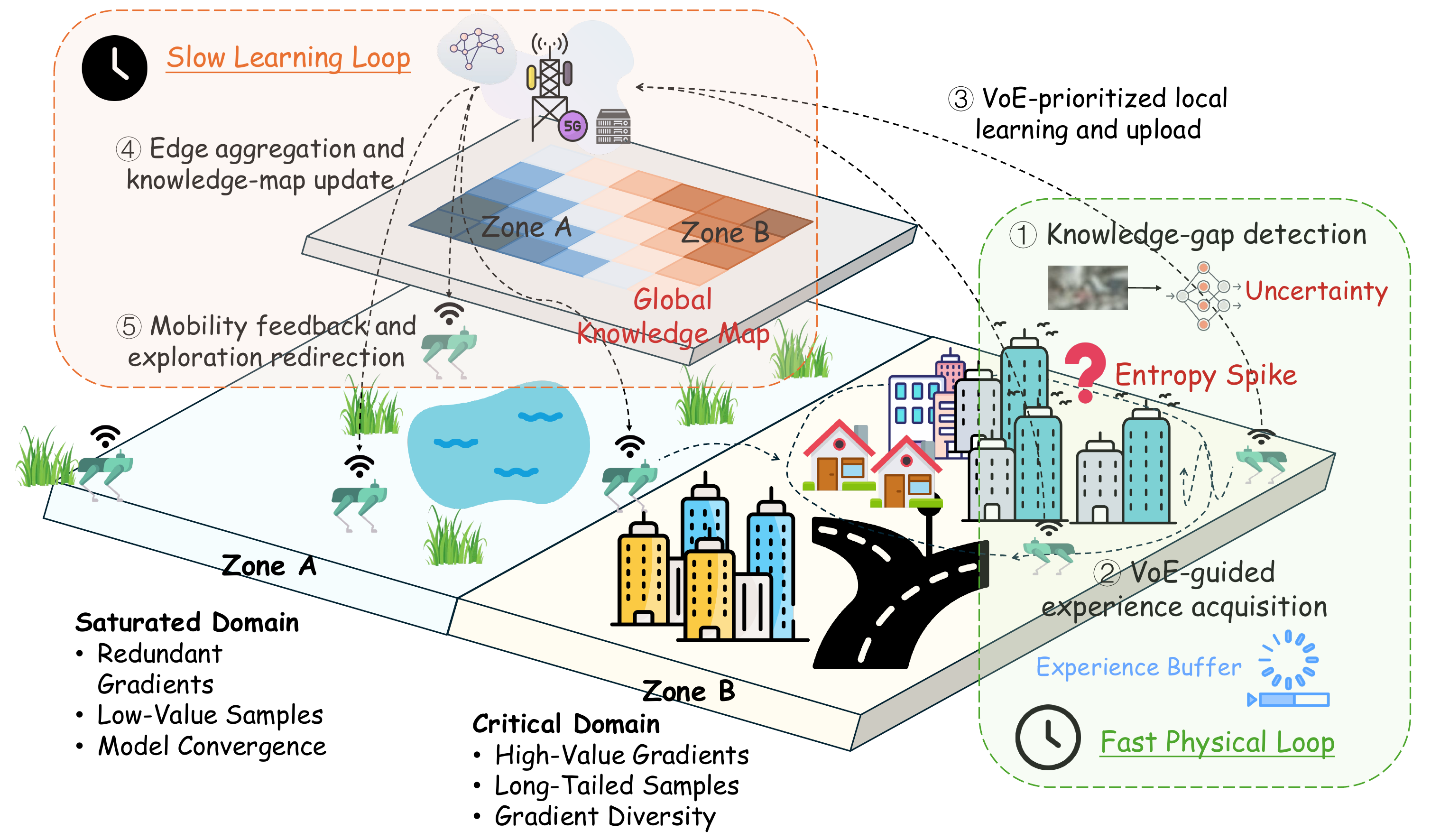}
	\caption{VoE-guided active embodied federated learning. Agents detect knowledge gaps, use expected VoE to guide physical experience acquisition, and upload high-value local updates, while the edge refreshes the global knowledge map for subsequent mobility guidance. }
	\label{fig:ei for edge}
\end{figure}

\subsection{Active Embodied Federated Learning}

Communication-oriented metrics such as age of information, age of incorrect information, and value of information mainly evaluate information that has already been generated \cite{yang2022semantic}. These metrics do not determine which physical action should be taken to generate more useful future experience.

ENEI therefore defines VoE as an action-conditioned measure of the expected learning utility of future embodied experience. VoE combines agent-side predictive uncertainty with an edge maintained regional gain estimate.
Uncertainty indicates a potential knowledge gap, whereas regional gain reflects how much previous experience from
similar regions improved the global model. The gain estimate is updated from observed learning outcomes and then used to guide
subsequent experience acquisition and update prioritization. This combination prevents communication resources from being repeatedly allocated to uncertain but unlearnable observations.
For example, persistent sensor noise may produce high uncertainty, but its regional gain decreases when previous uploads provide little improvement. Consequently, high VoE is assigned to experience that is both locally novel and historically
learnable.


VoE plays two complementary roles. At the experience level, it prioritizes acquired samples for local training and model updates for uplink transmission. At the action level, expected VoE guides
agents toward regions where valuable experience is more likely to emerge. The edge periodically feeds a global knowledge map, including regional mastery, historical gains, and remaining knowledge gaps, back to the agents. Each agent then combines this information with its local uncertainty, task objective, and movement cost to adjust its trajectory, sensing viewpoint, or interaction behavior.

Because physical interaction and federated optimization operate at different temporal scales, AEFL adopts a two-timescale
workflow. Agents continuously sense, act, and buffer experience, whereas local training and model upload are performed
periodically or when communication resources are available. Partial participation and asynchronous aggregation further accommodate intermittent connectivity and mobility-induced availability.

The AEFL workflow consists of four steps, as illustrated in Fig.~\ref{fig:ei for edge}.

\noindent \textbf{Step 1: Knowledge-gap detection.}
During task execution, each agent evaluates the predictive uncertainty of its local model and identifies the environmental regions or contexts associated with potential knowledge gaps. \textit{(Embodied device)}

\noindent \textbf{Step 2: VoE-guided experience acquisition.}
The agent combines local uncertainty with the regional gain estimates provided by the edge to estimate the future VoE of candidate actions. It then adjusts its mobility, sensing viewpoint, or interaction behavior toward regions where informative and learnable experience is more likely to emerge. \textit{(Embodied device)}

\noindent \textbf{Step 3: VoE-prioritized local learning and upload.}
The agent uses the acquired high-VoE experience for lightweight local training. Under limited communication and computing resources, agents with higher-value updates are prioritized for uplink transmission to the edge. \textit{(Embodied device)}

\noindent \textbf{Step 4: Edge aggregation and knowledge-map update.}
The edge aggregates the received model updates, evaluates their actual regional learning gains, and updates the global knowledge map accordingly. \textit{(Edge server)}

\noindent \textbf{Step 5: Mobility feedback and exploration redirection.}
Based on the updated knowledge map and regional learning-gain estimates, the edge provides mobility feedback to embodied agents, guiding subsequent exploration toward underrepresented and learnable regions while deprioritizing saturated or persistently ambiguous ones.  \textit{(Edge server and embodied devices)}

Through this closed loop, physical actions shape the experience distribution, VoE identifies which experience is worth communicating, and edge-side learning guides the next embodied actions. AEFL therefore extends conventional federated learning (FL) from passive aggregation over fixed local datasets to action-driven acquisition and communication of mobility-generated experience.

\subsection{Case Study: VoE-Guided AEFL for Mobility-Driven Edge Learning}

\begin{figure*}
	\centering
	\includegraphics[width=18cm]{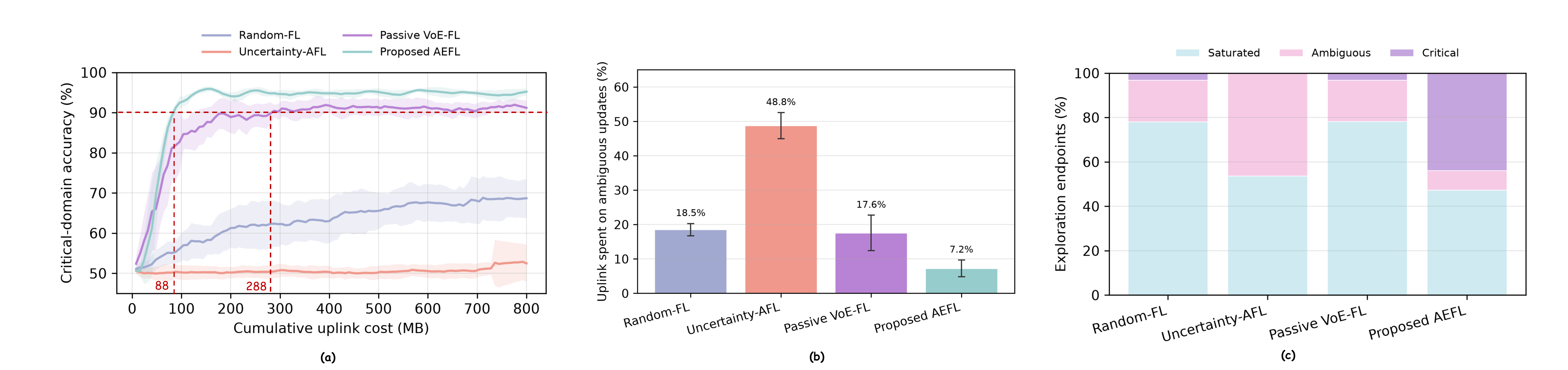}
	\caption{Performance evaluation of VoE-guided active embodied federated learning. (a) Communication-efficient edge learning: critical-domain accuracy versus cumulative uplink cost, showing that the proposed AEFL reaches the target accuracy with substantially lower communication overhead than Passive VoE-FL. (b) Ambiguous-update suppression: the proportion of uplink resources spent on persistently ambiguous updates, demonstrating that VoE reduces ineffective model transmissions compared with uncertainty-based prioritization. (c) Action-induced experience distribution: exploration endpoint distributions under different strategies, showing that expected VoE guides agents toward the learnable critical domain. Shaded regions and error bars indicate the variation over ten random seeds.}
	\label{fig:aefl}
\end{figure*}

To illustrate how embodied mobility can improve federated edge learning, we conduct a lightweight proof-of-concept simulation that couples experience acquisition, uplink prioritization, and edge model evolution. We consider a ten-position mobility corridor containing three domains. The \textit{saturated domain} follows the pretraining distribution and provides limited additional gain. The \textit{critical domain} introduces a learnable distribution shift through a $90^{\circ}$ feature rotation, while the \textit{ambiguous domain} contains high-entropy but randomly labeled observations that cannot be reliably learned.

Twelve agents start from a common waypoint and perform five movement actions before collecting local data. A tabular Q-learning policy controls mobility. The uncertainty-driven baseline uses predictive entropy as its movement reward, whereas AEFL uses expected VoE, combining local uncertainty with the regional gain estimated by the edge. Each agent then collects 32 samples and performs five local training steps. Under a fixed uplink budget, four agents upload 2~MB model updates in each FL round. The edge aggregates the updates, refreshes the regional gain estimates, and feeds the resulting knowledge map back to the agents. We compare Random-FL, Uncertainty-AFL, Passive VoE-FL with random mobility, and the proposed AEFL over 100 FL rounds and ten random seeds.

As shown in Fig.~\ref{fig:aefl}(a), AEFL reaches 90\% critical-domain accuracy with 88~MB of cumulative uplink traffic, compared with 288~MB for Passive VoE-FL, reducing communication cost by 69.4\%. After 100 rounds, their corresponding accuracies are 95.20\% and 91.18\%. Since both schemes use VoE-based update prioritization, this gap mainly reflects the benefit of using expected VoE to guide physical experience acquisition.

The resulting uplink efficiency is shown in Fig.~\ref{fig:aefl}(b). Uncertainty-AFL spends 48.78\% of its communication budget on ambiguous updates. Passive VoE-FL reduces this ratio to 17.58\% by learning that repeated updates from the ambiguous domain provide limited gain. AEFL further lowers the ratio to 7.25\%, because it not only suppresses low-value updates after collection, but also redirects agents away from regions unlikely to generate useful experience.

Fig.~\ref{fig:aefl}(c) shows that AEFL directs 43.87\% of exploration endpoints toward the learnable critical domain, whereas Passive VoE-FL reaches this domain in only 3.22\% of the cases. In contrast, Uncertainty-AFL directs 46.13\% of its explorations toward the ambiguous domain, demonstrating that uncertainty alone cannot distinguish learnable knowledge gaps from irreducible ambiguity.

Overall, this proof-of-concept study shows that the benefit of AEFL comes not only from prioritizing valuable updates after
experience has been collected, but also from using edge-side knowledge to shape where that experience is acquired. By guiding
agents toward learnable and underrepresented regions while deprioritizing persistently ambiguous ones, AEFL improves the quality of local training data, accelerates critical-domain learning, and uses limited uplink resources more effectively.


\section{Future Directions}\label{Future}
\subsection{Mutual Reasoning Alignment for Edge-Orchestrated Embodied Swarms}

ENEI introduces edge-assisted collaborative reasoning to promote compatible behaviors among heterogeneous embodied agents. A key future challenge is how to maintain such mutual reasoning alignment efficiently as the number, roles, and physical states of agents continuously change. Rather than continuously exchanging complete reasoning states, the wireless edge should identify and deliver compact agent-specific information that can correct belief divergence, coordinate policy responses, and preserve physical-action compatibility. This calls for communication-efficient alignment mechanisms that explicitly account for the downstream effects of transmitted information on distributed reasoning and joint behaviors.


\subsection{Lifelong Edge Learning from Mobility-Driven Experience}
Future frameworks should address how global models at the edge can continuously assimilate open-world experiential knowledge gathered by mobile agents without suffering from catastrophic forgetting. This demands macroscopic system designs for decentralized, long-term memory consolidation, and scalable model evolutionary mechanisms over distributed edge networks.

\subsection{Actuation-Centric Cross-Layer Design for 6G Networks}
To fully realize the potential of edge-embodied collaboration, next-generation wireless architectures must move toward a complete co-design of communication protocols, edge computing schedules, and physical control loops. Future studies should establish comprehensive network optimization frameworks that map multi-agent physical risks and mission criticalities directly to physical-layer communication resources. This shift will replace conventional throughput- or content-centric metrics with pragmatic, actuation-aware orchestration paradigms to ensure collective utility and operational safety of the entire swarm.

\section{Conclusion}

In this article, we introduced ENEI as an action-aware wireless edge framework that establishes a 6G-mediated bidirectional edge--embodiment loop. ENEI positions the edge as an action-aware cognitive service provider, while the 6G fabric serves as the programmable interaction medium for goal-oriented wireless delivery. The framework is organized along two reciprocal empowerment axes. The edge-to-embodiment axis preserves local autonomy through confidence-aware edge assistance and compact skill adaptation under changing environments, whereas the embodiment-to-edge axis enables agents to actively acquire valuable physical experience and feed mobility-driven knowledge back to evolve edge intelligence. The two case studies illustrate the feasibility and communication efficiency of these complementary mechanisms. Overall, ENEI provides a new perspective on wireless edge intelligence in which physical actions not only consume edge cognitive services, but also continuously ground and reshape edge intelligence through embodied interaction.



\bibliographystyle{IEEEtran}
\bibliography{IEEEabrv,myre}

\end{document}